\documentclass[pdflatex,sn-mathphys-num]{sn-jnl}

\usepackage{graphicx}%
\usepackage{multirow}%
\usepackage{amsmath,amssymb,amsfonts}%
\usepackage{amsthm}%
\usepackage{mathrsfs}%
\usepackage[title]{appendix}%
\usepackage{xcolor}%
\usepackage{textcomp}%
\usepackage{manyfoot}%
\usepackage{booktabs}%
\usepackage{algorithm}%
\usepackage{algorithmicx}%
\usepackage{algpseudocode}%
\usepackage{listings}%

\makeatletter
\renewcommand{\thefigure}{\arabic{figure}}
\renewcommand{\fnum@figure}{Supplementary Figure~\thefigure}
\makeatother

\makeatletter
\renewcommand{\thetable}{\arabic{table}}
\renewcommand{\fnum@table}{Supplementary Table~\thetable}
\makeatother

\theoremstyle{thmstyleone}%
\theoremstyle{thmstyletwo}%

\theoremstyle{thmstylethree}%

\usepackage{changes}
\definechangesauthor[name={reviewer 3}, color=orange]{R3}

\definechangesauthor[name={reviewer 2}, color=magenta]{R2}

\definechangesauthor[name={reviewer 1}, color=green]{R1}

\definechangesauthor[name={journal format}, color=gray]{J}

\begin{document}

\title[Verification of the two-fluid model for dilute two-phase flow.]{Experimental evidence for granular shear-flow instability in the Epstein regime}



\author*[1]{\fnm{Holly L.} \sur{Capelo}}\email{holly.capelo@unibe.ch}
\author[2,3]{\fnm{Jean-David} \sur{Bod\'enan}}
\author[1]{\fnm{Martin} \sur{Jutzi}}
\author[1]{\fnm{Jonas} \sur{K\"uhn}}
\author[2]{\fnm{Cl\'ement} \sur{Surville}}
\author[2]{\fnm{Lucio} \sur{Mayer}}
\author[3]{\fnm{Maria} \sur{Schönbächler}}
\author[1]{\fnm{Yann} \sur{Alibert}}
\author[1]{\fnm{Nicolas} \sur{Thomas}}
\author[1]{\fnm{Antoine} \sur{Pommerol}}

\affil*[1]{\orgdiv{Physics institute, Space Research and Planetary Sciences}, \orgname{University of Bern}, \orgaddress{\street{Sidlerstrasse 5}, \city{Bern}, \postcode{CH-3012}, \country{Switzerland}}}

\affil[2]{\orgdiv{Center for Theoretical Astrophysics and Cosmology, Institute for Computational Science}, \orgname{University of Zurich}, \orgaddress{\street{Winterthurerstrasse 190}, \city{Zurich}, \postcode{CH-8057}, \country{Switzerland}}}

\affil[3]{\orgdiv{Institute of Geochemistry and Petrology}, \orgname{ETH Zurich}, \orgaddress{\street{Clausiusstrasse 25}, \city{Zurich}, \postcode{8092}, \country{Switzerland}}}


\abstract{Stability analysis of two-fluid protoplanetary disc models has enriched our understanding of how solids can grow into larger bodies called planetesimals. Dust particles entrained in a gas stream modify the flow, creating shear layers prone to instability. In such environments, drag occurs in the free-molecular (Epstein) regime. Recreating these two-phase flows on Earth is difficult due to gravity-driven buoyancy. Here we use particle image velocimetry to study a low-pressure dust-gas mixture at Knudsen numbers up to 10 in microgravity. We observe a granular shear flow instability, characterized by a periodic velocity field, which can be modeled to first order as a Kelvin-Helmholtz (KH) instability. This behavior resembles a Kelvin-Helmholtz instability and provides a benchmark for two-fluid theories relevant to planet formation.}





\maketitle

\section{Introduction}

In models of gaseous circumstellar discs, dust grains sediment to form a dense layer of particles \cite{goldreich_ward}, or collect in filaments near local pressure maxima \cite{Surville:2018}. As these regions provide a contrast in dust and gas mass densities, the gas and solid phases experience resistance from one another, through collisions of dust particles with gas molecules and transfer of momentum back to the gas. This process of mutual solid-gas coupling creates regions with differential velocities, that result in particle-rich layers shearing against adjacent particle-poor gas layers \cite{Weidenschilling:1997,Weidenschilling:1980}. At the most fundamental level, this phenomenon can be understood by studying two shearing fluids of different density that have a velocity difference at their interface; under conditions where the two phases differ sufficiently in mass, density and velocity, perturbations are amplified enabling growth of waves and the onset of a fluid instability. Well-known examples of analogous processes under atmospheric conditions, occur where particulate matter such as cloud droplets or aerosols hover and drift with respect to the gas \cite{kh_atmosphere}. In these cases, the formation of periodic wave patterns and turbulence are important for determining the transport and concentration of particulate matter \cite{bonadonna}. It is therefore justified to assume that related flow processes can govern the distribution and transport of dust grains in protoplanetary discs \cite{Sekiya_2000,Gomez_2005,Johansen_2006,Bai_2010a}. 

Indeed, the particle-gas interaction is considered critical for driving disc evolution and converting primitive solids—including sub-$\mu$m size dust grains, icy particles, and cm-size pebbles—into planetesimals \cite{WeidenschillingCuzzi:1993,Johansen_2014}. There is debate about whether a thin layer of dust at the disc midplane might fragment due to self-gravity forming planetesimals or whether this process could be stalled if the dust layer density is limited by the turbulence generated by a Kelvin-Helmholtz-like shear instability (henceforth KH instability). It has also been pointed out that for enhanced dust densities, the relative motion between the dust and gas phases would lead to a runaway streaming instability which concentrates material sufficiently for planetesimals to form \cite{You_good2005}, and that this could compete effectively against diffusive turbulence \cite{Bai_2010a}. Other authors have proposed a general class of dust-drag driven instability with similar control parameters \cite{Squire_Hopkins:2018a}. This broad class of so-called `resonant drag instabilities’ (RDIs) arises from differential motion between the particle and gas phases. Although there is a large body of theoretical studies of unstable two-phase flow, experimental studies on this topic are very rare. To remedy this issue, \cite{lambrechts} showed that the signature effects of the streaming instability, in particular spontaneous particle clustering in a mass-loaded fluid, could be reproduced in a simplified non-rotating system, and provided useful scaling relationships that apply flexibly to either the sedimentation of dust particles in a protoplanetary disc, or the sedimentation of particles in a dilute laboratory flow. \cite{Capelo:2019} performed experiments in an equivalent system and confirmed spontaneous clustering dynamics on the appropriate scale, and also found evidence of a collective particle-drafting effect for particles in the Stokes drag regime. The current experimental work is not focused upon the streaming instability, but similarly involves a system that is greatly simplified to capture and isolate a fundamental mechanism that underlies theory. Important distinctions apply to the growth rate and critical length scales one expects for various types of instability, which cannot, and need not, all be reproduced in experiments.  Instead, this work places utmost importance upon testing the generally accepted premise that the solid-particle phase behaves as if it is a fluid. It is therefore the goal of this work to reproduce a standard fluid-dynamical phenomenon under conditions that have not been achieved before; specifically, we address the case of a potential KH-like shear flow instability, where one of the phases consists of solid dust-particle analogue material. Furthermore, it is necessary to test the analogy between atmospheric flows and the flow conditions found in circumstellar contexts, with consideration for the specific parameters and properties of the modeled fluid.

Treating free-floating dust particles collectively as a fluid simplifies the modelling of the particle-gas processes in the earliest stages of planet formation \cite{Nakagawa}. It is acceptable to consider the particle phase as a pressureless fluid, since the particle number density is low enough that collisions are rare \cite{Jacquet_2011}. The average dust-to-gas mass density ratio, $\epsilon \equiv $ $\rho_{dust}/\rho_{gas}$ governs the force acting in opposition between the two phases. The quantity $\rho_{dust}$ is the dust mass contained per unit volume, and derives from the number of particles per unit volume, the particles' average size and material density (see also definitions in the Subsection Flow Parameters). The quantity $\rho_{gas}$ is simply the gas density, assuming an ideal gas. In a circumstellar context the mean free path of the gas $l$ is typically equal to, or larger than, the size of the particles $d_{p}$, resulting in a high Knudsen number $Kn \equiv l/d_{p}$. Therefore the applicable drag force $F_{\rm d}$ corresponds to either transitional or free-molecular flow -- also referred to as Epstein drag \cite{Weidenschilling:1977a}. Inviscid flow (Euler) equations apply when inertial effects dominate viscous forces within a fluid. The Reynold's number compares the relative strengths of these forces: $Re = F_{inertial}/F_{viscous} = \rho_{\rm gas} v L/\mu$. Gas velocity is $v$, the length scale is $L$ and the molecular viscosity is $\mu$. For a large $L$, the inertia is driven up and internal fluid friction is therefore neglected. For particles in a flow, the particles' momentum competes with the fluid drag force resulting in a characteristic time for particle acceleration to slow down to a steady velocity, known as the particle friction timescale, $T_{f} = m \delta v/F_d$, for a particle of mass $m$ and differential velocity $\delta v$.  The parameter $T_{f}$ determines whether particles directly follow fluid motions, or else obey their own inertia. A dimensionless Stokes number is typically defined by comparing $T_{f}$ with a system timescale \cite{Armitage_book}. It has been shown that moderately decoupled solids are effective at driving fluid instabilities both in rotating systems—such as the streaming instability—and in non-rotating, local settings, where the so-called settling or drafting instability applies due to the negligible role of Coriolis forces at the particle friction length scale \citep{lambrechts,Capelo:2019}. In addition, RDIs can also develop in non-rotating environments through coupling to acoustic or buoyancy modes \cite{Squire_Hopkins:2018a}. For high enough dust to gas mas density ratio $\epsilon\sim1$, a streaming instability will result. However, shearing-type instabilities can result even for very moderate values of $\epsilon \lesssim 0.1$, corresponding to small, tightly coupled particles \cite{Surville:2018}. There are known instabilities that are driven even in systems without rotation, and by comparing the theory of such instabilities with the present experiment, the validity of the two-fluid approximation may be confirmed.
 
Models that rely heavily upon dust-gas processes to enable planetesimal formation and feed the planetary core accretion process can produce systems that are consistent with those that we observe \citep{Lambrechts:2019, Liu:2020}. However, there is currently no way to directly observe the two-fluid hydrodynamics using astronomical data. In discs, there is expected to exist a differential, or `drift' velocity between the phases, which results from sedimentation, or due to the difference between Keplerian dust velocities and pressure-supported sub-Keplerian gas. Thus far, deviations from Keplerian rotation, particularly
those arising from pressure support, are difficult to detect with current observational precision \cite{rosenfeld_2012}. 
In a controlled setting, we can recreate this differential motion to study how the two phases interact directly, without external forces such as gravity. In the absence of external body forces, we expect that the dust and gas phases approach an equilibrium velocity due directly to the mutual drag forces. Otherwise, when external forces are present, distinguishing between the effects of these forces and the intrinsic dust-gas interactions becomes challenging.
In particular, we test the hypothesis that in the reference frame between the particles and gas, a moderately mass-loaded fluid with high $Kn$ can cause a strong enough back-reaction effect to sufficiently slow the flow and result in a shear instability. To test this hypothesis, it is necessary to avoid gravitational sedimentation and introduce particles directly into a low-pressure gas stream. To this purpose, we designed and built the Timed Epstein Multi-Pressure Vessel at Low Accelerations (TEMPus VoLA \cite{Capelo:2022}), which was dedicated to testing for the existence of a granular shear instability in high Knudsen-number flow. 

The underlying objective of this work has been to substantiate the two-fluid model that is the standard theoretical framework for treating dust dynamics in protoplanetary discs. We investigated the potential presence of a shear-flow instability, since its existence implies coupling between the momentum of the dust and gas phases. The experiment involves the shearing interaction between a particle layer and a laminar fluid flow ($Re_{c}\sim7$ on the container scale and $10^{-5}\lesssim Re_{p}\lesssim 10^{-4}$). KH instabilities are characterized by wave-like structures that emerge at the interface between two layers of fluid (or between a fluid and particle layer in this case) due to velocity shear.  Our analysis of the flow-velocity field shows persistent periodic behaviour. Further analysis of the spectrum of frequencies present in the flow field revealed a complex system with emergent pattern formation. These oscillations are present across multiple time steps, suggesting that the system is not purely laminar but instead exhibiting dynamic instability. A purely laminar flow would not be expected to generate such wave patterns, but the presence of a shear layer due to the particle-fluid interaction provides a natural environment for the development of KH instabilities.

The growth of structure in the flow is highly suggestive of the nonlinear phase of a two-fluid instability that arises naturally without external forces applied. This discovery provides direct empirical evidence for the type of dynamics that are predicted to spontaneously occur near the midplanes or in pressure bumps in protoplanetary discs. Verifying the fluid-like behavior of the particle phase establishes a foundation for further studies in higher-quality, longer-term microgravity experiments. These future studies will enable high-precision investigations of dust hydrodynamics in fully-developed, unstable flows involving rarefied gas.

\section{Results}\label{sec:results}
\subsection{Two-fluid system} \label{subsec:two_fluid_model}
The system of study involves a parallel Pouissele flow in a cylindrical container at submbar gas pressure. 10-$\mu m$ silicon dioxide particles are introduced at a steady rate along the center line of the flow. There are two stages that need to be considered with regards to the system dynamics: first, the momentum coupling between the dust and gas phases that should generate a velocity shear profile with a mass density difference; secondly, the fluid instability that would naturally result from this density and velocity difference. We later present evidence for the existence of the complex flow dynamics associated with the instability and rule out other explanations for this behavior. The development of complex flow dynamics is conditional upon the particle phase behaving like a pressureless fluid that acts on the gas in the first phase of the setup. Thus, we will consider the measurements of unstable flow to be valuable evidence toward the correctness of the two-fluid model approach itself.

\begin{figure}[h!]
\includegraphics[width=12cm]{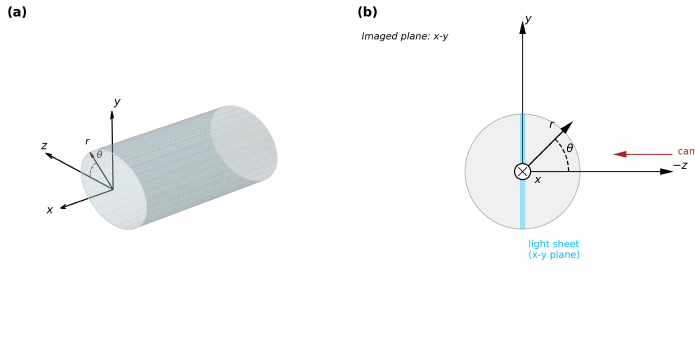}
\centering
\caption{ \label{fig:coordinates} Coordinate system used in this work. {\bf(a)}: overlay of cylindrical and Cartesian coordinates onto cylindrical pipe flow. The $r-\theta$ plane corresponds to the cylinder cross section. The direction of flow in this system is in the positive x-direction.  {\bf(b)}: cut-away view of the cylinder cross section, indicating the imaging plane with respect to the camera placement.}
\end{figure}

\begin{figure}
\includegraphics[width=12cm]{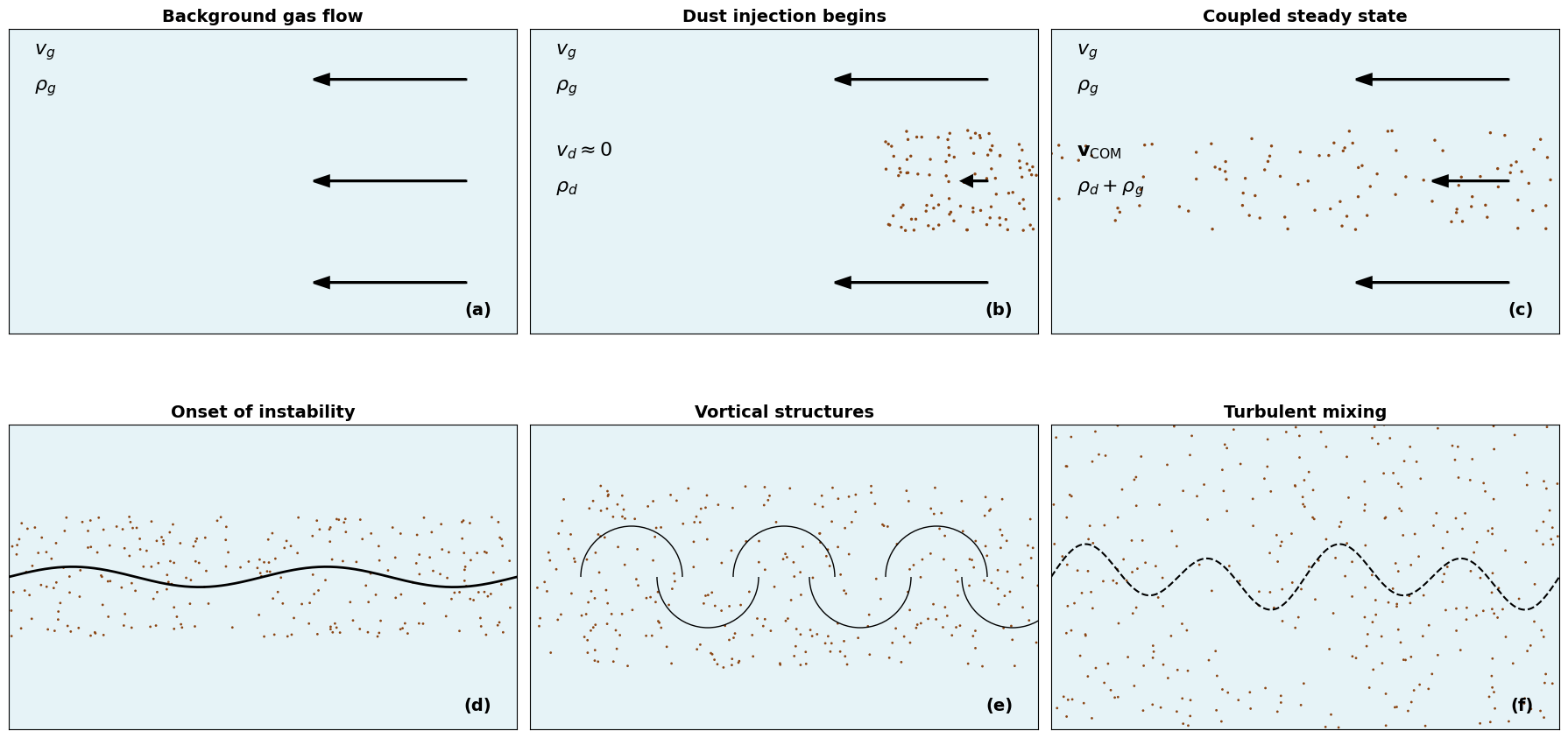}
\centering
\caption{ \label{fig:cartoon}Basic principle of the system, shown as a slice on the $x-r$ plane. (a): the background laminar gas flow has uniform velocity;
(b): dust particles start from rest when injection begins; (c): as the dust and gas act upon one another at the flow's midline, the mixture assumes a slower velocity than the original gas speed i.e. the center of mass dust-gas velocity, $v_{\rm com} < v_g$.; (d): it is expected that the presence of a differential velocity and density will lead to the onset of an instability; (e): as the instability develops, vortices may form; (f): in a fully developed state, the dust and gas will be mixed by turbulence.} 
\end{figure}

Figure \ref{fig:coordinates} illustrates the geometry of the flow. We assign the x-coordinate direction to the long axis of the cylinder and refer to this as the axial direction. The figure also illustrates the light sheet that is pulsed in synchronicity with a high-speed camera and records the particle positions, to enable the determination of a velocity field, using the particle image velocimetry (PIV) technique. The illuminated slice corresponds to the x-y plane. The radial and azimuthal directions are then the projections onto the y-z plane which defines the cross-section of the cylinder. For a cylindrical pipe flow with constant mass flux rate, the gas velocity and pressure would be constant at all radii. In contrast, in this system, dust particles start from rest and enter the flow by being dragged by the gas. Two-fluid models predict that, due to the momentum coupling between dust and gas, the particles and gas in the midline should come to rest with respect to one another in their center of mass frame; the dust-laden gas layer will then have a density and velocity different from those of the original gas flow, which continues to have its original velocity and density above and below the midline. Figure \ref{fig:cartoon} illustrates the basic principle of the system.

The Supplementary Note 1, Supplementary Equations 1-8, provides the three-dimensional two-fluid equations in the coordinate systems shown in Figure \ref{fig:coordinates}. The dynamical equations include the spatially averaged dust and gas mass densities, $\rho_{\rm p}$ and $\rho_{\rm g}$, respectively. The gas velocities denoted by $u$ and dust velocities by $v$, with directional subscrpts corresponding to the coordinate system in Figure \ref{fig:coordinates}. There are several simplifications that allow us to greatly reduce the model prescription for the dust-gas mixture that originates in the midline of the flow. Consider that i) The gas is isothermal; ii) There is no density gradient in either the gas or particles, since there is no hydrostatic equilibrium condition in the absence of gravity; iii) The laminar gas flow travels only in the axial direction and transports the particles in the axial direction. Note also that the linear gas velocity through the pipe with fixed cross section is constant (see \cite{Capelo:2022} and the Section Methods of this work). Therefore we can neglect the pressure gradient in the gas momentum equations, which are fully derived in Supplementary Note 1. Moreover, we can assume axisymmetry about the cylinder center and that all velocities are strictly axial (\( u_x, v_x \)), and functions of \( r \) and \( t \) only. The gas density is assumed to be constant.
Gas momentum in the axial direction is:
\begin{align}\label{equn:gas_momentum}
\frac{\partial u_x}{\partial t}
= \frac{1}{t_{\rm f}} \frac{\rho_{\rm p}}{\rho_{\rm g}} (v_x - u_x).
\end{align}
The Particle density is constant on average in time at each location.
And the particle momentum is:
\begin{align}\label{equn:particle_momentum}
\frac{\partial v_x}{\partial t}
= -\frac{1}{t_{\rm f}} (v_x - u_x).
\end{align}
Recall that the friction time, $t_{\rm f}$, is the characteristic timescale over which the particles approach their terminal velocity, with respect to the gas velocity.  
In this system, individual dust particles accelerated from rest would be expected to travel at their terminal velocity in the direction of the gas flow. In the absence of external forces, the terminal velocity of the particles would be the same as the laminar background gas velocity, e.g. the velocity shown in panel (a) of Figure \ref{fig:cartoon}. More formally, the equation of motion for individual particles would be $v_x(t) = u_x \left(1 - e^{-t/t_f}\right)$. In this system, for a laminar flow and not considering the collective dynamics of the particle population, one would expect the particles to simply have constant $v_x=v_g$. Concretely, for $ T_{f}$ of 3–6 ms and gas flow speed of 1 m/s, the particles would couple to the gas speed in the first 9–18 ms after they start from rest and before they travel $\approx$ 6–12 mm of distance. Therefore individual particles have time to reach terminal velocity before passing through the measurement window. See also the Section Methods for details of the particle properties, flow speed, and measurement window placement.

 Instead, if the particle phase behaves collectively as a massive fluid acting upon the gas, it is expected that the dust and gas in the dust layer will approach a common speed. This can be seen from inspection of equations \ref{equn:gas_momentum} and \ref{equn:particle_momentum}, where the velocity of the gas reaches an equilibrium velocity with the gas, that is equal in the center-of-mass frame of reference, and differs from the original background gas flow velocity; see e.g. panel (c) Figure \ref{fig:cartoon}. See also Figure 19 from \cite{Capelo:2022}. We emphasize that the approach to conditions where the dust and gas velocities are equal only applies to the particle layer at the midline. The gas flow closer to the chamber walls should not respond to the dust.  The dust-filled gas is therefore of a different density and velocity than the rest of the gas flow, as depicted in Figure \ref{fig:cartoon}. In this model, it is the average dust density that determines the dynamics and this is set by both the thickness of the dust layer and the seeding density. The initial dust layer is determined by the size of the vacuum chamber inlet which is less than 1/4 of the diameter of the entire flow vessel diameter of 8 cm. The particles can diffuse due to noise and--in the event of an instability-- due to the increasingly complex velocity field.
 
As the system approaches a steady state, we can consider the dust-filled layer as a fluid with a unique velocity and density. In this formulation, we consider the layer at the midline to be effectively a fluid of mixed dust and gas, which assumes a single velocity and density, represented as $v_2$ and density, $\rho_{2}$ = $ (1+\epsilon)\rho_{g}$. The gas that is not initially acted upon by the dust retains the original values $\rho_{1}$ = $\rho_{g}$ and $v_{1}=v_{g}$.  Therefore, collective dust-gas coupling would result in a shear profile, which can become unstable. The instability is significant, insofar as it can only result from the dynamic coupling between dust and gas, and therefore indicates that the two-fluid approach is valid. We therefore proceed to check the PIV data for the existence of a shear instability to verify that the dust and gas have become dynamically coupled, as predicted.

The general scenario described above and represented in figure \ref{fig:cartoon} can be seen intuitively in the raw imaging data see data availability statement. At first, the dust seeding is very sparse, corresponding to a low dust-to-gas density ratio, $\epsilon$. As time progresses, the particle dynamics becomes increasingly complex and the dust density in the field of view increases significantly. The dynamical complexity necessitates a quantitative analysis of the velocity vectors assigned to the particles in the imaging data. 

\subsection{Pattern formation}

Inspecting the time evolution of the particle velocity fields, a pattern is immediately noticeable, whereby some regions apparently possess very little movement, and these regions appear in alternation above and below the midline of the field of view. These regions also apparently contain non-zero curl, particularly as the steady flow conditions persist past the initial particle seeding stages. The divergence of the velocity field was calculated as $\nabla \cdot \vec{v} = \frac{\partial u}{\partial x} + \frac{\partial v}{\partial y}$, where $u$ and $v$ are the velocity components in the $x$ and $y$ directions. This was achieved using numerical differentiation of the velocity fields. For exemplary snapshots of the two-dimensional velocity fields, vorticity and divergence calculations, see Supplementary Figures 1,2, and 3. We also provide time-evolved statistics on the average component-wise velocities and divergence thereof, see Supplementary Figure 4. We notice that despite localized variations in these quantities, the averages in the measurement volume over time remain close to zero. This shows us already that the incompressibility condition of the flow, $\nabla \cdot \vec{v} = 0$, is met on the `global' scale at which we observe. It is also immediately clear that the data is not in agreement with any expectations from particles coupled to a laminar gas flow, for which the particles would simply be moving uniformly in the positive x-direction at their expected relaxation velocity of $\simeq 0.8 $ m s$^{-1}$.  
\subsection{Velocity line profiles}
\begin{figure}
\includegraphics[width=15 cm,trim=0cm 8cm 0cm 0cm, clip]{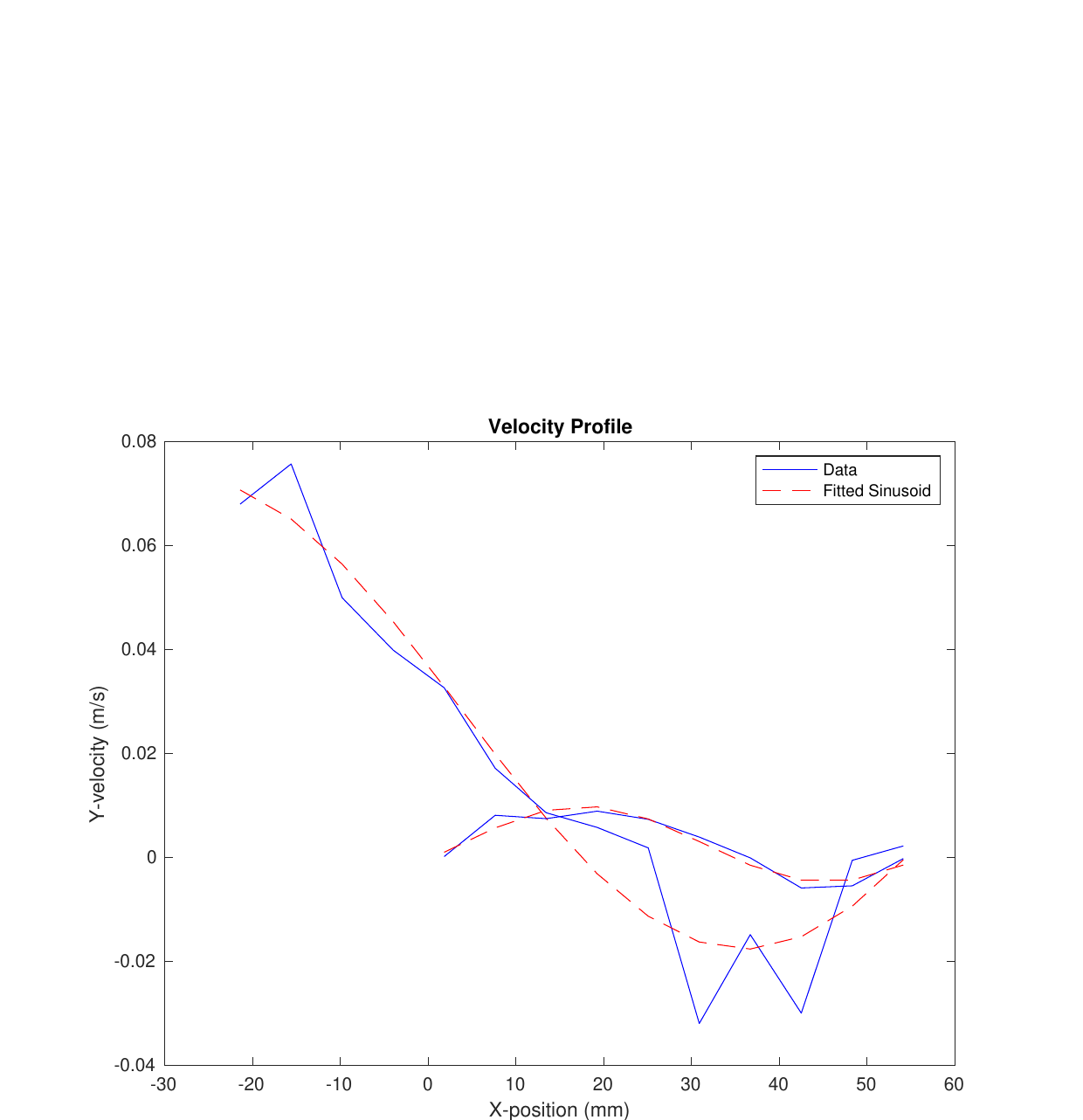}
\centering
\caption{ \label{fig:wave_fits}Line profiles of the velocity in the y-direction, taken during the apparent onset phase of the fluid instability. Solid blue lines are the extracted data and the dashed red lines represent sinusoidal fits to the data. Each profile derives from a single image in the time series, for clarity only a few exemplary fits are shown. }
\end{figure}

We address the flow dynamics during the stage where the flow is developing periodicity and significant vorticity, shown in panel (d) of figure \ref{fig:cartoon}.
To simplify the analysis of the complex velocity field, we extract line profiles for all frames from $t$ = 4 s to $t$  = 22 s of our measurement, holding the y-coordinate fixed at the midline, and reporting the vertical velocity across all values of the horizontal axis, i.e. the direction of shear. In Figure \ref{fig:wave_fits}, we illustrate this procedure with examples of the line profiles extracted from two frames, and sinusoidal fits to the data.  
As can be seen in Figure \ref{fig:wave_fits}, the vertical velocity at the mid-line of the flow varies from positive to negative values. Moreover, from one frame to the next, representing a time difference of 1 ms, wave-like velocity patterns possess different amplitude and period properties. We only show two examples of the full distribution of extracted line profiles. In the next sub-section, we address the full spectrum of oscillation frequencies over time. We then utilize statistics of all extracted midplane velocity line profiles, to estimate the characteristic minimum length-scale of the flow features.
\subsection{Wavelet analysis}

\begin{figure}
\includegraphics[width=14 cm]{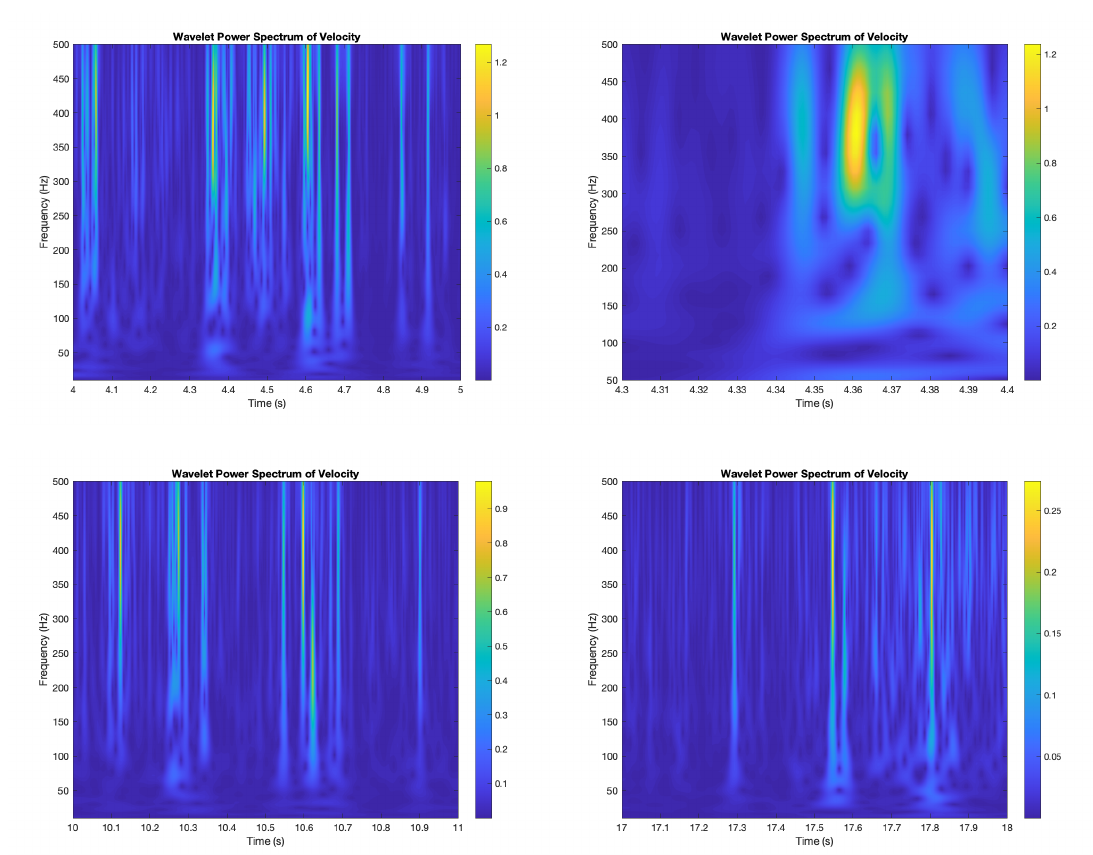}
\caption{\label{fig:power_spectrum} Wavelet frequency spectrum at select time intervals. From  $t$ = 4 s to $t$ = 5 s of the measurement, the correlation with the wave function is maximal. The dominant frequencies are concentrated above 300 Hz and the correlation with the wave function is very strong. There is a very strong feature occurring around $t$ = 4.36 s, shown in closeup.  At two later times, $t$ = 10 s to $t$ = 11 s and $t$ = 17 s to $t$ = 18 s, the range of frequencies gradually tends towards lower values and the magnitude of the coefficients decreases over time.}
\end{figure}

We perform an analysis using the Morlet wavelet to examine the time-frequency spectrum of oscillatory behavior \citep{Percival_Walden_2000}; details of the methodology are provided in the Section Methods. The wavelet coefficients result in the power spectrum shown in Figure \ref{fig:power_spectrum}. The x-axis represents time, the y-axis represents the frequency (related to the scale \( a \)), and the color intensity represents the magnitude of the wavelet coefficients. We show examples of the spectrum at the times in the measurement sequence: 4-5s, 10-11s, and 17-18s. A close-up view of the spectrum from 4.3-4.4s allows inspection of where the range of frequencies is concentrated. The upper limit on the figures at 500 Hz, corresponds to the Nyquist frequency, i.e. half of the 1KHz sampling rate.    

The vertical stripes in the wavelet power spectrum indicate periodic or quasi-periodic oscillations that recur consistently over time. These stripes represent increased wavelet power at certain frequencies that remain relatively stable or repeat at intervals. The intervals are not perfectly constant, but correspond roughly to a time separation on the order of $\sim$0.01 s. This time scale is consistent with the gas advection speed, which moves across the $\sim 8cm$ measurement window at 0.8-1 m s$^{-1}$. Most likely, these stripes are the result because the structures at hierarchical scales in the flow travel past the measurement window at the advection speed. However, we suggest that these features that appear at different scales and frequencies may propagate partially in the azimuthal direction as well. However, we cannot decompose the different directions of travel and possible signatures on the spectrum with our two-dimensional measurements.  Of note is the decrease in the range of the coefficients over time, as well as the gradual decrease in magnitude. For instance, in second 4 of the measurement, the maximum magnitude of the coefficient shown by the color bar is 1.8, and decreases to 0.9 in second 10, and to 0.25 in second 17. Moreover, the features in the power spectrum with the highest coefficients gradually shift to lower frequencies.

The evolution of the power spectrum, together with the length scales associated with the fluctuations in the velocity field, indicates that the initial stages are more characterized by high-frequency wave-like behavior, which gradually develops into more complex flow with larger features.  However, we note that we do not observe a typical transition through a linear growth phase that emerges out of an originally laminar flow.  
The measurements, which occur at an upstream point in the flow, strongly indicate that the system has transitioned into a pattern-forming state.
To estimate the critical wavelength of the instability, we analyzed the velocity profiles during the early, approximately linear stage of growth, selecting data starting from $t$ = 4s of the measurement sequence. We applied sinusoidal fits to the transverse velocity profiles and found that the smallest consistently observed wavelength, as well as the mode of the resulting wavelength distribution, was approximately 3 cm. We interpret this as the critical wavelength of the instability, defined as the shortest wavelength that exhibits sustained growth during the linear regime -- see panel (c) in figure \ref{fig:cartoon} and thus marks the smallest spatial scale at which coherent structures are expected to emerge.

 We find this to agree in order of magnitude with the predictions from the linear stability analysis of the expected fluid instability \cite{Capelo:2022}, which predicts the fastest growing wavelengths are those with frequencies greater than approximately 100 Hz at wavelengths below the cm length scale. Note that these predictions are made with the simplifying assumption that the direction of shear dominates the dynamics, and that the amplification of wave-like perturbations is occurring primarily in the projected two-dimensional plane. In the Section Conclusions below, we discuss potential limitations to this two-dimensional assumption.  

\subsection{KH dispersion relation}

First, we recall that the dispersion relation for an oscillatory instability consists of a wave phase velocity and expected growth rate \cite{chandrasekhar1961,Pringle:2007tn}, which we will denote $v_{\text{phase}}$ and $\sigma$, respectively. In this section, we use the measured critical wavelength and range of observed frequencies to determine the model parameters which would produce the oscillatory instability. We will confirm that the requisite parameters are consistent with the prior knowledge of our system. 

 The gas and dust velocities were computed by numerically solving the discrete form of equations \ref{equn:gas_momentum} and \ref{equn:particle_momentum}:
\begin{equation}
v_{\text{gas}}(t_{i+1}) = v_{\text{gas}}(t_i) + \frac{\epsilon}{t_f} \left( v_{\text{dust}}(t_i) - v_{\text{gas}}(t_i) \right) \Delta t 
\end{equation}
 and 
 \begin{equation}v_{\text{dust}}(t_{i+1}) = v_{\text{dust}}(t_i) + \frac{1}{t_f} \left( v_{\text{gas}}(t_i) - v_{\text{dust}}(t_i) \right) \Delta t,
 \end{equation}
 
 where $v_{\text{gas}}(t_0)$ is initialized at $1 \, \text{m s$^-1$}$, $v_{\text{dust}}(t_0)$ is initialized at $0 \, \text{m s$^-1$}$, and $\Delta t$ is the numerical time step. The steady-state velocity of the dust, $v_2$, was determined by iterating these equations over time until convergence. From the steady state condition, we can consider the dust layer, composed of dust an gas with common center of mass velocity, to have density, $\rho_{2}$ = $ (1+\epsilon)\rho_{g}$. The gas that is not initially acted upon by the dust simply has $\rho_{1}$ = $\rho_{g}$.

The phase velocity $v_{\text{phase}}$ was calculated:
\begin{equation}
v_{\text{phase}} = \frac{\rho_1 v_1 + \rho_2 v_2}{\rho_1 + \rho_2},   
\end{equation} where $v_1$ and $v_2$ are the steady-state velocities of the gas and dust phases, respectively.

The predicted frequency corresponding to the measured 3 cm wavelength was then calculated as 
\begin{equation}
f = \frac{v_{\text{phase}}}{\lambda},
\end{equation} where $\lambda = 0.03 \, \text{m}$. These frequencies were compared with the experimentally observed frequency range of $300-500 \, \text{Hz}$. Only combinations of $\epsilon$ and $t_f$ that yielded predicted frequencies within this range were considered valid.

Since we do not measure exactly at the dust injection point, but instead downstream, we use the model to check the likely initial dust-to-gas ratio  $\epsilon$. We define a dimensionless parameter $\sigma \cdot t_{\text{travel}}$, where $\sigma$ is the theoretical KH growth rate and $t_{\text{travel}}$ is the particle travel time to the measurement window. The parameter $\sigma \cdot t_{\text{travel}}$ quantifies how much the instability has grown during the transit of particles to the measurement location. The theoretical growth rate, $\sigma$, is given by the dispersion relation of the KH instability:
\begin{equation}
\sigma = k_x \cdot (v_1 - v_2) \cdot \frac{\sqrt{\rho_1 \rho_2}}{\rho_1 + \rho_2},
\end{equation} where $k_x = 2\pi / \lambda$ is the wavenumber, $\lambda = 3 \, \text{cm}$ is the measured wavelength, $v_1$ and $v_2$ are the steady-state velocities of the gas and dust phases, respectively, and $\rho_1 = \rho_g$ and $\rho_{2}$ = $ (1+\epsilon)\rho_{g}$ are the gas and dust densities.

The condition $\sigma \cdot t_{\text{travel}} > 1$ indicates that the instability has grown significantly by the time the particles reach the measurement window, whereas $\sigma \cdot t_{\text{travel}} < 1$ implies that the growth is incomplete. Since the fluid instability observed in the experiment appears to have progressed past the linear onset stage (evidenced by significant vorticity development in the measurement window), we constrained $\epsilon$ to values where $\sigma \cdot t_{\text{travel}} \leq 1$. To calculate $t_{\text{travel}}$, we used the relationship $t_{\text{travel}} = d / v_2$, where $d = 0.1 \, \text{m}$ is the distance to the measurement window, and $v_2$ is the steady-state dust velocity obtained by solving the coupled momentum equations described above. By evaluating the relationship between $\sigma \cdot t_{\text{travel}}$ and $\epsilon$ (with $t_f = 0.006 \, \text{s}$ fixed), we identified a maximum value of $\epsilon$ where $\sigma \cdot t_{\text{travel}} = 1$, which is 0.065 and corresponds to steady-state dust velocity v$_{2}$ = 0.939 m s$^{-1}$
and estimated travel time of 0.320 seconds. This additional constraint on $\epsilon$ reconciles the observed instability dynamics with theoretical predictions of growth and transport. On the other hand, we have very little information about the emerging velocity field in the first 4 seconds of our measurement, because at such necessarily low mass loading the particle field is too sparse to assign a spatially averaged velocity field to the flow. Therefore we cannot rule out that the development time is longer. However, to be consistent with the frequency constraints above, the maximum $\epsilon$ can be is 0.175, resulting in a marginally different steady-state dust velocity and v$_{2}$ = 0.851 m s$^{-1}$ and estimated travel time of 0.352 seconds. Please see Supplementary Figure 5 for the visual representation of the relationship between $\sigma \cdot t_{\text{travel}}$ and $\epsilon$.

\section{Discussion}\label{sec:discussion}

The wavelet analysis that we performed upon the velocity line profiles confirmed that the measured oscillatory frequencies are consistent with the desired initial conditions which would lead to a shear layer with fastest growing features on the scale of those we observe. Moreover, these implied initial conditions are consistent with our design principles and a priori knowledge of our experimental conditions, such as the gas flow speed and particle types. 

We extracted the characteristic length scale by fitting sinusoids to the line profiles of the velocity field at the middle of the dust layer. However, it is to be noted that we did not directly observe the initial condition, as the seeding and transport of the initial dust layer to the observation point and the onset of the non-linear mode coupling apparently co-evolve on similar time scales.  

Nevertheless, we found that the instability does not appear to saturate within the measurement time of 24 seconds, as the coefficients in the wavelet power spectrum continue to drop over time, indicating an increasingly complex dynamic, as would be expected from the growth of the initial perturbations. On the other hand, we do not appear to be observing the fully developed turbulent state, in which case the mode coupling would be so strong that the velocity field would become highly irregular and the periodicity would likely disappear. We can therefore conclude that we have measured close to the onset phase and that longer times would be required to observe the fully-developed state.

Infinitesimal perturbations to the pressure and density of a homogeneous and isotropic medium should decay, unless there is a wave-number selection process, which amplifies the amplitude at a given frequency resulting in wave-like features. We find that, indeed, the particle phase can act upon the gas to bring about conditions for instability, in a fluid that would otherwise remain homogeneous in the absence of particles. The experimental conditions that led to discovering the pattern formation process have never before been achieved. Previous studies of multi-particle dynamics in dilute flows achieved transition flow conditions with $Kn \sim1$ \cite{Capelo:2019}, and somewhat related studies \cite{Schneider_2019} achieved only moderate $Kn \sim .01$, corresponding to continuum flow. In the two cases mentioned, however, linear drag laws could at least be assumed due to relatively low pressures and consequently low Reynolds numbers. The microgravity environment enables the extremely dilute two-fluid system, for which we neither need to address sedimentation dynamics, nor do we require high intrinsic fluid viscosity in order to suspend the particles. The unique conditions are ideal to test both the two-fluid equations, or the Einstein hypothesis \citep{Einstein:1905}. By eliminating external body forces, we confirm that the only source of momentum to excite the flow dynamics derives internally from the particle-gas drag coupling, which is assumed by the two-fluid approach that is the standard treatment for dust in protoplanetary discs.  Moreover, the dust particles are well separated, such that neither collisions nor long-range hydrodynamic interactions are expected to excite the particle velocities. Therefore, the fluid meets additional criteria which make the analogy to astrophysical flow correct.

Demonstrating the existence of a KH-like instability for particles in the Epstein regime has direct relevance for the potential to generate mid-plane turbulence and stir the sedimented dust layer in planet-forming discs. The results to a large extent confirm the expectation that such instabilities should occur when the conditions are met. 

Dust-drag type fluid instabilities are mainly studied for their ability to concentrate particles and facilitate the formation of planetesimals. It has also been noted that the dust is itself a driver of turbulence and vorticity. This is significant, because it is as of yet poorly understood how turbulence is generated in planet-forming discs. In particular, an effective turbulent viscosity is parameterized in this context, as it is needed for momentum transport in order to drive gas accretion. However, the mechanistic driver for the turbulence remains under discussion. The upper limit of turbulence intensity driven by dust motion in protoplanetary disks is estimated from the perspective of energetics by \cite{Takeuchi:2012}. Our experimental observation of a granular shear instability offers concrete evidence that dust itself can seed and sustain velocity fluctuations in a stratified, laminar gas flow. This supports the notion that dust-loading may be a direct source of turbulence, rather than merely a passive participant in existing instabilities. Such a mechanism could contribute to the anomalous angular momentum transport typically attributed to turbulent viscosity in accretion models. By demonstrating that shear instabilities can emerge from dust-gas interactions in the absence of externally imposed turbulence, our results provide a physical basis for dust-driven momentum transport and offer a candidate mechanism to account for the origin of turbulence in quiescent regions of planet-forming discs.

 \section{Conclusions}\label{sec:conclusions}
These experiments provide a foundation for a series of studies that aim to bridge theoretical models and empirical observations of dust-gas interactions. Several numerical codes now implement dusty fluid dynamics, and the availability of experimental data enables direct benchmarking of these simulations. The standard two-fluid model remains broadly applicable due to its simplicity, with the primary control parameters being the dust-to-gas mass ratio, $\epsilon$, and the particle stopping time, $t_{\rm f}$. By converting the experimental conditions into these quantities, we find that the general predictions of two-fluid momentum coupling are supported. However, the interpretation of collective particle behavior based only upon models is limited by the volume-averaged nature of $\epsilon$, which may obscure local fluctuations, spatial inhomogeneities, and clustering phenomena. Importantly, direct measurements can offer access to micro-physical quantities that are often inaccessible in simulations, including the relative velocity distribution of particles, the emergence of hierarchical structures, and collective drag effects across different flow regimes.

To enable a rigorous comparison between experiment and simulation, it would be preferable to specify not only the macroscopic control parameters but also the system geometry; see, for example, the full system of equations in Supplementary Note 1, Supplementary Equations 1-8. Additional features such as aspect ratio, boundary conditions, and confinement may influence higher-order mode couplings, which could play a role in the development and interpretation of observed instabilities. On the other hand, we have been careful to minimize the direct impact of the container. First, the flow chamber is big enough that particle interactions with the walls, generally important over a distance of a few particle diameters, can be neglected; second, we measure less than three pipe diameters downstream of the gas entry point, and so we avoid the development of a parabolic Hagen-Pouseille flow profile, which generally results from no-slip boundary conditions. Therefore, it should be sufficient to compare the experimental results with a numerical set-up that implements simplifications such as periodic boundary conditions. 

 Vibrations at the dust injection point, could indeed influence the thickness of the initial particle layer and hence add uncertainty to the initial dust-to-gas density ratio, $\epsilon$. As we have not directly visualized the dust injection, we do not know exactly how confined the original dust layer was. We recommend experiments in higher-quality micro-g conditions and calibration of the dust layer thickness to better control and verify the initial condition of the experiment. 

 In this work, we focused upon `structure' in the velocity field and made the assumption that the length scales associated with velocity fluctuations represent the length scales associated with the average dust-density variation. While this was sufficient to identify unstable flow, we are not able with the 2-D PIV method to study, for example, correlations in the fluctuations of density and velocity on a localized scale. We checked the divergence of the velocity as it is proportional to the changing rate of the density, $\partial \rho/ \partial t$. We did not pursue this analysis further because the 2-d measurement approach is not sure to lead to unambiguous results or interpretation. For example, in regions where the velocity tends to zero, we cannot determine the difference between whether there are no particles in those regions, if the particles are slowing down in the center of vorticies, or alternatively, if there is travel of the particles in the azimuthal direction (into or out of the page), which projected onto the X-Y plane would appear as zero velocity, and therefore also affect the calculation of the density based upon the divergence. Methods which track individual particles in 3D, usually referred to as Lagrangian Particle Tracking, would allow more in-depth analysis of spatial variation and corrections between velocity and density and other interesting statistics such as relative particle velocities in high-density regions. Such detailed analysis would allow to confirm how instabilities with dust comprising one fluid phase differ in detail from the more simple case of classical fluids and to test whether numerical simulations are correctly capturing such differences. 

We measured the pattern formation phenomena at a stage that is close to onset and have not yet studied the fully-developed state. We recommend further experiments to follow the full development, which would require longer time in microgravity. In order to witness all of the transitions from stable, to linear growth, to pattern forming, and finally turbulent states, many repetitions of the experiment should be run where the particle size and dust injection rate is carefully controlled and systematically varied so that the growth can be intentionally driven or suppressed.  We note that the conditions for a shear instability such as the one that we have generated here involve relatively low friction times and dust-to gas density ratios. However, as these quantities increase, it is expected that eventually a streaming-like instability should set in. Since the flow under consideration is sub-sonic, it is not expected that an acoustic RDI, of the type discussed in \cite{Squire_Hopkins:2018a, magnan:2024} is operating. However, there could be the potential to generate a non-linear drafting instability (streaming-like) of the type studied in the works of \cite{lambrechts} and \cite{Capelo:2019}, or a sub-sonic RDI discussed by \cite{Hopkins_Squire:2018}. Again, systematically varying the dust properties and dust injection rate, would allow a determination of the threshold in critical values where the streaming-like instability starts to set in.

\section{Methods}\label{sec:methods}

\subsection{Concept}
The TEMPusVoLA shear flow chamber was designed to satisfy the pre-requisites for flow instability to occur, while matching the set of dimensionless fluid parameters that would correspond to the flow regime that is typical of planet-forming discs. The apparatus is integrated into a secure rack that is suitable for human operators to use during parabolic flight. 

Despite the specific context-related properties of astrophysical flow, the granular shear-flow instability investigated here is a standard example of a system that is expected to follow a linear onset stage, provided that i) the system is axis-symmetric about the cylinder center, ii) it is isotropic in at least one dimension, and iii) the equilibrium state is steady. The descriptions that follow explain how the experiment meets these conditions. 

\begin{figure}
\includegraphics[width=12 cm]{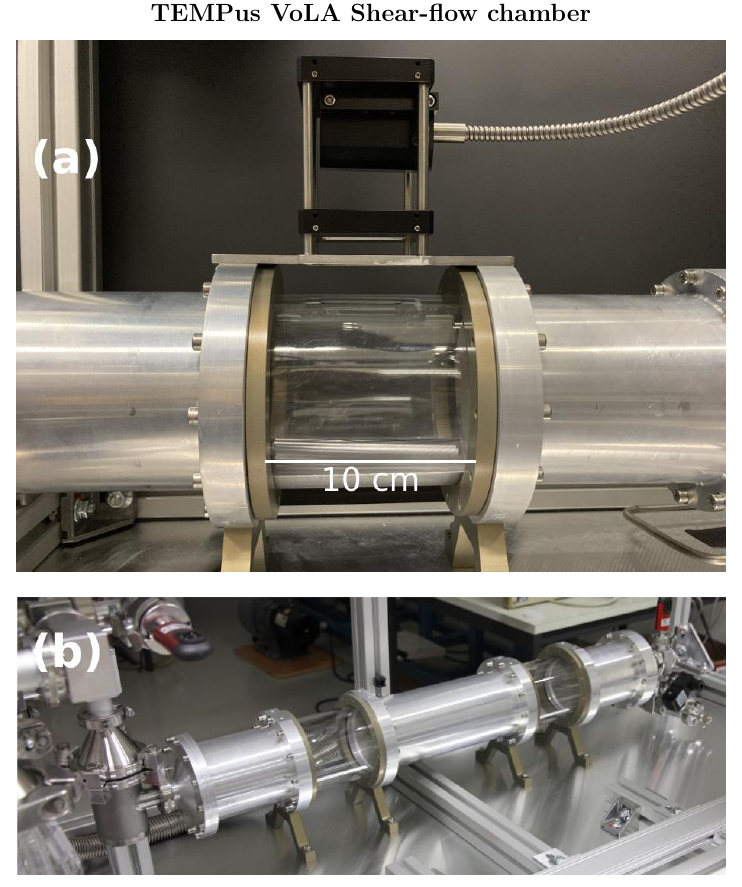}
\centering
\caption{\label{fig:apparatus} {\bf (a):} light-sheet optic mounted on top of the viewing port of the shear-flow chamber of the TEMPus VoLA microgravity experiment \citep{Capelo:2022}. {\bf (b):} The entire shear-flow chamber with dimensions $\sim1$ m and inner tube diameter 8 cm.  }
\end{figure}

First, the flow vessel is cylindrical, to satisfy the symmetry requirement. The sealed vessel contains a flow of air that is driven by a pressure gradient. A vacuum line is connected upstream, and a mass flow controller is installed downstream, to control the steady low-pressure conditions. We show a photograph of the shear-flow chamber in Figure \ref{fig:apparatus}

\begin{figure}
\includegraphics[width=6 cm]{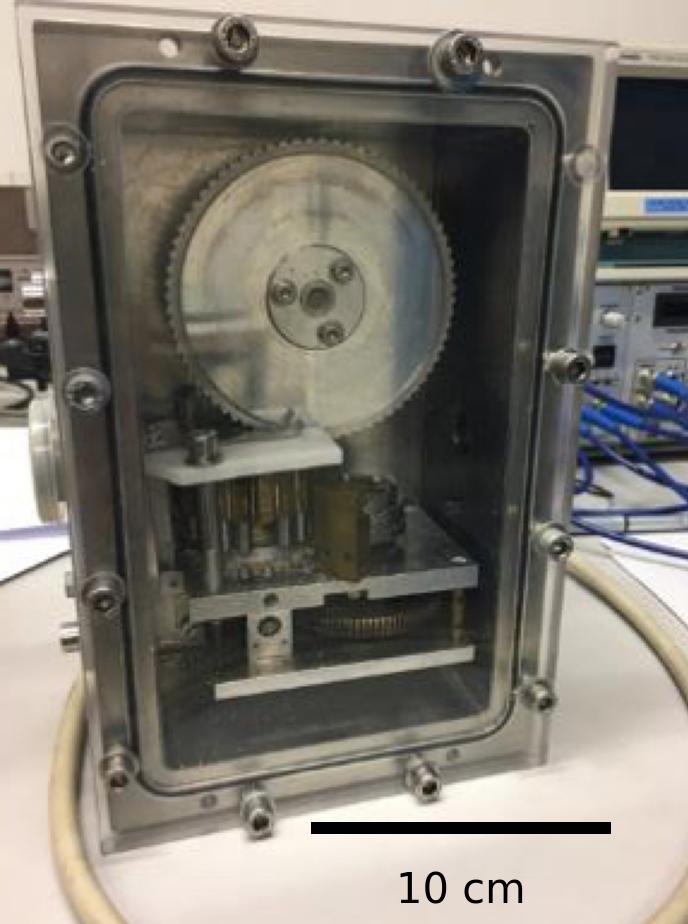}
\centering
\caption{\label{fig:dust_injector} The dust injector mechanism utilized in the experiments. A pre-loaded revolver holds dust material which comes into contact with a cog wheel that disperses the dust into the vacuum line.}
\end{figure}

We produce a stream of solid material with a dust injector that introduces particles into the midline of the flow at a continuous, steady rate, resulting in a particle distribution that is statistically isotropic in the transverse plane—that is, approximately uniform and without preferred direction on average. In figure \ref{fig:dust_injector} we show a photograph of the injector mechanism. The particles are loosely coupled to the gas, and follow the bulk velocity of the flow, due to the neutralised gravitational load in a parabolic flight environment. The zero-g phase of the parabola lasts approximately 23 seconds.   
\cite{Capelo:2022} presents calculations of the time for the system to reach a steady-state velocity once the particles are injected from rest. Even for low values of $\epsilon$, the inertial coupling between the two phases should cause the fluid speed to relax. We calculated that the equilibrium velocity caused by the back-reaction force of the particles should be reached within the first few seconds of dust transport. The optical observation window is placed accordingly, to measure the velocity field at a point in the flow where the steady-state should have been reached. In this work we will address the complicating feature of the system, which is that the approach to a steady-state velocity and the growth of the instability occur at the same time.  

\subsection{Measurement techniques}
Gas pressure was continuously monitored throughout the experiment using a Pfeiffer Vacuum pressure transducer, which was read out at a rate of 1Hz. This provided a stable reference for the background pressure in the chamber during dust injection and shear flow evolution. The resultant values are used to calculate the flow parameters are discussed below in Subsection Flow parameters.

Particle Image Velocimetry (PIV) is a non-intrusive optical measurement technique used to derive velocity fields in fluid flows. In this method, the flow is seeded with tracer particles, and the instantaneous velocity field is computed by correlating the spatial patterns of particle positions in successive image pairs. This provides a time-resolved, quantitative measurement of the flow dynamics across the entire field of view. In this work we do not work with perfect gas tracers, but we apply the technique to weakly inertial particles to study the dynamics of the particle phase. PIV measurements in this experiment were acquired using a high-speed imaging system and processed with the LaVision DaVis software suite. The system captured two-dimensional image sequences at a sampling rate of 1kHz, from which velocity vectors were computed using standard cross-correlation techniques. Please also see a representative shapshot of the vector field, and resultant derived quantities such as vorticity and divergence, in Supplementary Figures 1, 2, and 3.

\subsection{Flow parameters}\label{subsec:flow parameters}
The set of flow parameters in our experiment is summarized in Table 1. We have tuned the gas pressure and flow velocity as well as the particle size and seeding density to result in a system with Epstein particle drag, moderate particle coupling strength and values of $\epsilon$ much less than 1. The Knudsen number $Kn$ is used to assess the regime of flow.  We calculate $l$ to remain fixed at 92 +/- 0.3 micrometers. For the dust particles in the size range 1-10 micrometers, $Kn$  is therefore always greater than 9 for the largest particles and increases by a factor 10 for the smallest. Thus, all of the particles are experiencing free-molecular flow. However, on the container scale, with diameter 8 cm, $Kn$ remains at 0.0011, and so it is completely justified to consider the bulk flow as a continuum.  The container scale $Re \sim 7$ and the particle scale $Re \sim 10^{-4} - 10^{-5}$. At neither scale we expect the spontaneous generation of turbulence, as this generally requires $Re\sim 1000$. 

To assess the initial average dust-to-gas mass density ratio $\epsilon$, we first calculated the particle number density by analyzing a series of images. Using automated detection, we found an average particle count of $\bar{N} = 24.91$ particles per frame, with a standard deviation $\sigma_N = 3.42$ particles per frame. To convert number density to mass density, we estimated the volume and mass of each particle, modeled as ellipsoidal platelets with an aspect ratio of 5:1. The particle size range was 5–10 microns, so the semi-major axis $b$ was 2.5–5 microns, and the semi-minor axis $a = b/5$. Each particle's volume was calculated as $V_{\text{particle}} = \frac{4}{3} \pi a b^2$ and, using the material density of 3950 kg/m³, each particle’s mass was $m_{\text{particle}} = V_{\text{particle}} \cdot \rho_{\text{material}}$, yielding masses from $2.58 \times 10^{-14}$ to $2.06 \times 10^{-13}$ kg. 

The average particle mass density in the illumination volume ($V_{\text{illumination}} = 1.777 \times 10^{-5} \, \text{m}^3$) was then calculated as $\rho_{\text{particles}} = \frac{\bar{N} \cdot m_{\text{particle}}}{V_{\text{illumination}}} \approx 0.17 \pm 0.06 \, \text{kg/m}^3$, where uncertainties from particle size and count were propagated using the formula $\sigma_{\rho} = \sqrt{\left( \frac{\partial \rho}{\partial N} \sigma_N \right)^2 + \left( \frac{\partial \rho}{\partial V_{\text{particle}}} \sigma_{V_{\text{particle}}} \right)^2}$, where $\sigma_N$ is the standard deviation of the particle count and $\sigma_{V_{\text{particle}}}$ reflects the uncertainty in particle volume due to the size range. Finally, to determine the particle-to-gas mass density ratio, we used the ideal gas law to estimate air density at 0.9 mbar. Assuming a pressure $P = 90 \, \text{Pa}$, a molar mass of air $M = 0.029 \, \text{kg/mol}$, the gas constant $R = 8.314 \, \text{J/(mol K)}$, and a temperature $T = 298 \, \text{K}$, we calculated the air density as $\rho_{\text{air}} = \frac{PM}{RT} \approx 6.53 \times 10^{-2} \, \text{kg/m}^3$. This yielded a particle-to-gas mass density ratio of $\frac{\rho_{\text{particles}}}{\rho_{\text{air}}} \approx 2.6 \pm 0.9$. However, we note that the measurement window is placed 30 cm downstream of the dust injection point. As the particles are seeded into flow, the fluid instability is likely co-evolving with the velocity-relaxation between the two flow phases. In the Section Results, we combine our measurements with theoretical expectations of the wave phase velocity and growth rate to further constrain the initial conditions on the dust to gas ratio $\epsilon$. 

The particle coupling timescale, or friction time $t_f$, was calculated directly in the Epstein regime for particles with a platelet shape. The friction time is defined as $t_f = \frac{l \cdot \rho_p \cdot d_{\text{eff}}}{4 \cdot \mu \cdot S}$, where $l$ is the mean free path of the gas molecules, $\rho_p$ is the particle density, $d_{\text{eff}}$ is the effective diameter of the platelets, $\mu = 1.8 \cdot 10^{-5} \, \text{Pa·s}$ is the dynamic viscosity of air, and $S = 1.5$ is the shape parameter for platelets. The effective diameter, $d_{\text{eff}}$, was calculated as $d_{\text{eff}} = \sqrt{d_{\text{long}} \cdot d_{\text{short}}}$, where $d_{\text{long}}$ is the major axis and $d_{\text{short}}$ is the minor axis. This expression is adapted from the Epstein drag law \cite{epstein}, with shape corrections following formulations used in studies of non-spherical particle dynamics \citep{BAGHERI2016526}. The shape parameter $S$ accounts for the departure of the particle shape from a perfect sphere. For platelets with an aspect ratio of $5:1$, $S = 1.5$ was used, which reflects the increased drag associated with this geometry relative to a sphere of the same effective diameter. This adjustment is necessary to ensure the Epstein drag law properly accounts for the particle's non-spherical nature. For particles with $d_{\text{long}} = 5 \, \mu\text{m}$ and $d_{\text{short}} = 1 \, \mu\text{m}$, $t_f$ was calculated to be $t_f = 4.535 \cdot 10^{-3} \, \text{s}$. For particles with $d_{\text{long}} = 10 \, \mu\text{m}$ and $d_{\text{short}} = 2 \, \mu\text{m}$, $t_f$ was calculated to be $t_f = 9.070 \cdot 10^{-3} \, \text{s}$. These calculations provide the coupling timescale for particles with different effective diameters and the assumed shape parameter in the Epstein regime. The analysis in the Section Results shows consistency with this range of friction times. These values are simultaneously consistent with the a priori estimates and modeled constrains upon the value of $\epsilon$.

\begin{table}[h]
\centering
\begin{tabular}{|c|c|}
\hline
\textbf{Parameter} & \textbf{Experimental Value}  \\
\hline
Re$_{\rm p}$ & 10$^{-5}$ - 10$^{-4}$   \\
\hline
$Kn_{\rm p}$ &$\geq$ 9    \\
\hline
T$_{\rm f}$ & 0.006-0.009 s   \\
\hline
 $\epsilon$ & 0.06-0.17   \\
\hline
Re$_{\rm c}$  & 7   \\
\hline
$Kn_{\rm c}$  &  0.0011  \\
\hline
\end{tabular}
\caption{Parameters and experimental conditions. The subscripts p and c indicate  Reynolds number and Knudsen number calculated on the particle scale or container scale, respectively. We report the particle friction time, mean dust-to-gas density ratio, as well as the mean pressure, and mean particle velocity in the direction of shear. }
\label{tab:example}
\end{table}
\subsection{Oscillation analysis}
To evaluate the spectrum of oscillatory frequencies present in the particle phase of the flow, and evaluate changes over time, we perform a wavelet analysis for the entire set of images. The continuous wavelet transform (CWT) is a mathematical technique used to decompose a signal into time-frequency space, enabling the identification of both the frequency content and its temporal localization. The wavelet transform of a signal \( x(t) \) with respect to a chosen wavelet function \( \psi(t) \) is defined by the following integral: \( W(a, b) = \int_{-\infty}^{\infty} x(t) \, \psi^*\left(\frac{t - b}{a}\right) \, dt \). In this equation, \( W(a, b) \) represents the wavelet coefficient at a given scale \( a \) and translation \( b \), where:
 \( a \) is the scale parameter, which is inversely related to frequency. Smaller values of \( a \) correspond to higher frequencies, while larger values of \( a \) correspond to lower frequencies.
 \( b \) is the translation parameter, which corresponds to a shift in time.
 \( \psi(t) \) is the mother wavelet, a localized oscillatory function that is stretched or compressed depending on the value of \( a \).
 \( \psi^*(t) \) denotes the complex conjugate of the wavelet function.
 The integral performs a convolution between the signal \( x(t) \) and the scaled and translated version of the wavelet.

We applied the Morlet wavelet for the continuous wavelet transform, which consists of a sine wave modulated by a Gaussian envelope, ensuring both time and frequency localization. The scale parameter \( a \) controls the dilation of the wavelet, allowing us to capture oscillations at different frequencies within the signal. The resulting wavelet coefficients, \( W(a, b) \), represent the strength of the signal's oscillatory behavior at different scales and times. Large coefficients indicate that the signal contains strong oscillations at that particular scale and time, while smaller coefficients suggest weaker or absent oscillations.

 To be mentioned, is that the parabolic flight environment is noisy in both the sense that the aircraft cabin can have loud noises, and in the sense that variations to the acceleration at the level of milli-g result from the atmospheric turbulence that the air-craft experiences. However, at the gas pressures in our experimental system, the sound speed is $\sim 345$ m s$^{-1}$, which is much greater than the wave phase velocities that we measure. So the waves that we observe cannot be merely particles responding to the pressure perturbations induced by sound waves. Regarding vibrations, we can neither explain the observed oscillations by shaking of the camera or of the flow vessel. Such artifact would manifest as the entire flow field moving up and down or side to side over subsequent frames, rather than produce localized sub-structures in the flow field such as those that we observe. 

\subsection{Procedures}

The operation and measurement sequence is coordinated with the aircraft maneuvers. A single parabola consists of separate phases: steady flight corresponding to Earth's gravity, $g$; the `pull-up' and `pull-out' phases corresponding to hypergravity which is $\sim 1.8 g$; and the microgravity, $0g$ phase. We operate the vacuum line continuously through all stages so that the gas pressure is simply fixed.  We also obtain uninterrupted pressure measurements from a transducer placed immediately upstream of the chamber, and log the output with one-second binning. We wait until the entry into the $0g$ phase to trigger both the dust injector and the high-speed cameras. The dust stream flows continuously throughout the entire $0g$ phase and we use high-speed cameras to acquire images over this entire time frame. We deduce the particle velocity field using a standard PIV technique. In order to access a slice of the velocity field at the chamber midline, a light sheet is installed parallel with the long axis of the cylinder. The data reduction and signal processing is described fully in \cite{Capelo:2022}.

\backmatter

\section{Data availability}
The datasets generated during and analysed during the current study are available in the \cite{zenodo} repository, with doi:10.5281/zenodo.10853363.
\bmhead{Supplementary information}

\bmhead{Acknowledgements}
H.L.C. and M.J. acknowledge support from the Swiss National Science Foundation (project number 200021\_207359). This work has been carried out within the framework of the NCCR PlanetS supported by the Swiss National Science Foundation under grants 51NF40\_182901 and 51NF40\_205606. We acknowledge the ongoing support of ESA, Prodex, and the Swiss Space Office. We thank the three anonymous referees for their careful review and constructive suggestions, which have improved the quality of this paper.

\bmhead{Author Contributions}
H.L. Capelo led the manuscript writing and performed the data analysis. Data processing was carried out by H.L. Capelo and J.-D. Bodénan. Measurement campaigns were conducted by H.L. Capelo, J.-D. Bodénan, A. Pommerol and J. Kühn. The instrument was designed and commissioned by H.L. Capelo, A. Pommerol and J. Kühn. \added{Interpretation of the scientific results and conceptualization of the project were contributed by H.L. Capelo, J.-D. Bodénan, M. Jutzi, J. Kühn, C. Surville, L. Mayer, M. Schönbächler, Y. Alibert, N. Thomas and A. Pommerol.} Supervision was provided by N. Thomas and A. Pommerol. Funding acquisition was undertaken by A. Pommerol, H.L. Capelo, M. Jutzi, J. Kühn, L. Mayer, M. Schönbächler, Y. Alibert and N. Thomas.

\bmhead{Competing Interests}
All authors declare no competing interests.

\bibliography{shearing}

\clearpage
\section*{Supplementary Note 1: Two-fluid equations in cylindrical coordinates}\label{appendix:two-fluid-equns}
The model has 8 parameters ($\rho_{\rm p}, \rho_{\rm g}, v_x,v_y, v_z,
u_x, u_y, u_z$) and the same number of equations. These are the spatially averaged dust and gas mass densities, with gas velocities denoted by $u$ and dust velocities by $v$. External forces acting on the fluid, such as gravity, can be neglected because we work in zero-g. In our system, it is meaningful to convert to cylindrical coordinates. We consider the axial direction to be along \( x \), with radial and azimuthal directions \( r \) and \( \theta \), respectively.

The gas continuity equation is:
\begin{align}
\frac{\partial u_x}{\partial x} + \frac{1}{r} \frac{\partial}{\partial r} (r u_r) + \frac{1}{r} \frac{\partial u_\theta}{\partial \theta} + \frac{1}{r} \frac{\partial}{\partial \theta} (\rho_{\rm g} v_\theta) = 0
\label{eq:cont_g_cyl}
\end{align}
and the gas momentum equations:
\begin{align} \label{equn:full_gas_momentum_cylindrical}
\text{Axial (}x\text{):} \quad 
&\frac{\partial u_x}{\partial t}
+ u_x \frac{\partial u_x}{\partial x}
+ u_r \frac{\partial u_x}{\partial r}
+ \frac{u_\theta}{r} \frac{\partial u_x}{\partial \theta}
= -\frac{1}{\rho_{\rm g}} \frac{\partial P}{\partial x}
+ \frac{1}{t_{\rm f}} \frac{\rho_{\rm p}}{\rho_{\rm g}} (v_x - u_x) \\
\text{Radial (}r\text{):} \quad 
&\frac{\partial u_r}{\partial t}
+ u_x \frac{\partial u_r}{\partial x}
+ u_r \frac{\partial u_r}{\partial r}
+ \frac{u_\theta}{r} \frac{\partial u_r}{\partial \theta}
- \frac{u_\theta^2}{r}
= -\frac{1}{\rho_{\rm g}} \frac{\partial P}{\partial r}
+ \frac{1}{t_{\rm f}} \frac{\rho_{\rm p}}{\rho_{\rm g}} (v_r - u_r) \\
\text{Azimuthal (}\theta\text{):} \quad 
&\frac{\partial u_\theta}{\partial t}
+ u_x \frac{\partial u_\theta}{\partial x}
+ u_r \frac{\partial u_\theta}{\partial r}
+ \frac{u_\theta}{r} \frac{\partial u_\theta}{\partial \theta}
+ \frac{u_r u_\theta}{r}
= -\frac{1}{\rho_{\rm g} r} \frac{\partial P}{\partial \theta}
+ \frac{1}{t_{\rm f}} \frac{\rho_{\rm p}}{\rho_{\rm g}} (v_\theta - u_\theta)
\end{align}
The particle continuity equation is:
\begin{align}
\frac{\partial \rho_{\rm p}}{\partial t}
+ \frac{\partial}{\partial x} (\rho_{\rm p} v_x)
+ \frac{1}{r} \frac{\partial}{\partial r} (r \rho_{\rm p} v_r)
+ \frac{1}{r} \frac{\partial}{\partial \theta} (\rho_{\rm p} v_\theta) = 0
\label{eq:cont_p_cyl}
\end{align}
and the particle momentum equations:
\begin{align}\label{equn:full_particle_momentum_cylindrical}
\text{Axial:} \quad 
&\frac{\partial v_x}{\partial t}
+ v_x \frac{\partial v_x}{\partial x}
+ v_r \frac{\partial v_x}{\partial r}
+ \frac{v_\theta}{r} \frac{\partial v_x}{\partial \theta}
= -\frac{1}{t_{\rm f}} (v_x - u_x) \\
\text{Radial:} \quad 
&\frac{\partial v_r}{\partial t}
+ v_x \frac{\partial v_r}{\partial x}
+ v_r \frac{\partial v_r}{\partial r}
+ \frac{v_\theta}{r} \frac{\partial v_r}{\partial \theta}
- \frac{v_\theta^2}{r}
= - \frac{1}{t_{\rm f}} (v_r - u_r) \\
\text{Azimuthal:} \quad 
&\frac{\partial v_\theta}{\partial t}
+ v_x \frac{\partial v_\theta}{\partial x}
+ v_r \frac{\partial v_\theta}{\partial r}
+ \frac{v_\theta}{r} \frac{\partial v_\theta}{\partial \theta}
+ \frac{v_r v_\theta}{r}
= -\frac{1}{t_{\rm f}} (v_\theta - u_\theta)
\end{align}

Examples of the velocity flow field, and derived quantities such as divergence and curl can be found in Supplementary Figures 1-3. These figures correspond to a single time stamp in the measurement sequence. For the full time series, of velocity vector field, please see data at Supplementary Reference 1.  
\begin{figure}
    \centering
    \includegraphics[width=15cm]{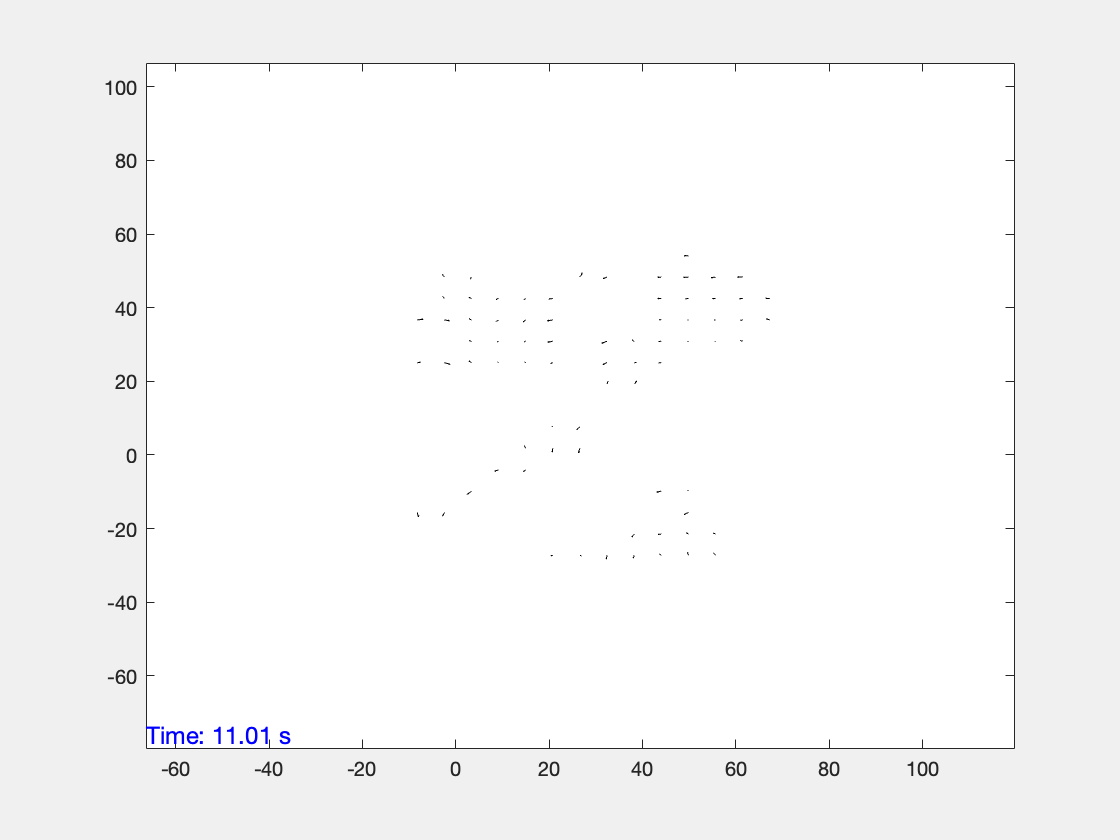}
    \caption{\label{fig:vfield} Example of the particle velocity field at the beginning of $t = 11$ s. Axis labels are in units of mm.}
\end{figure}

\begin{figure}
    \centering
    \includegraphics[width=15cm]{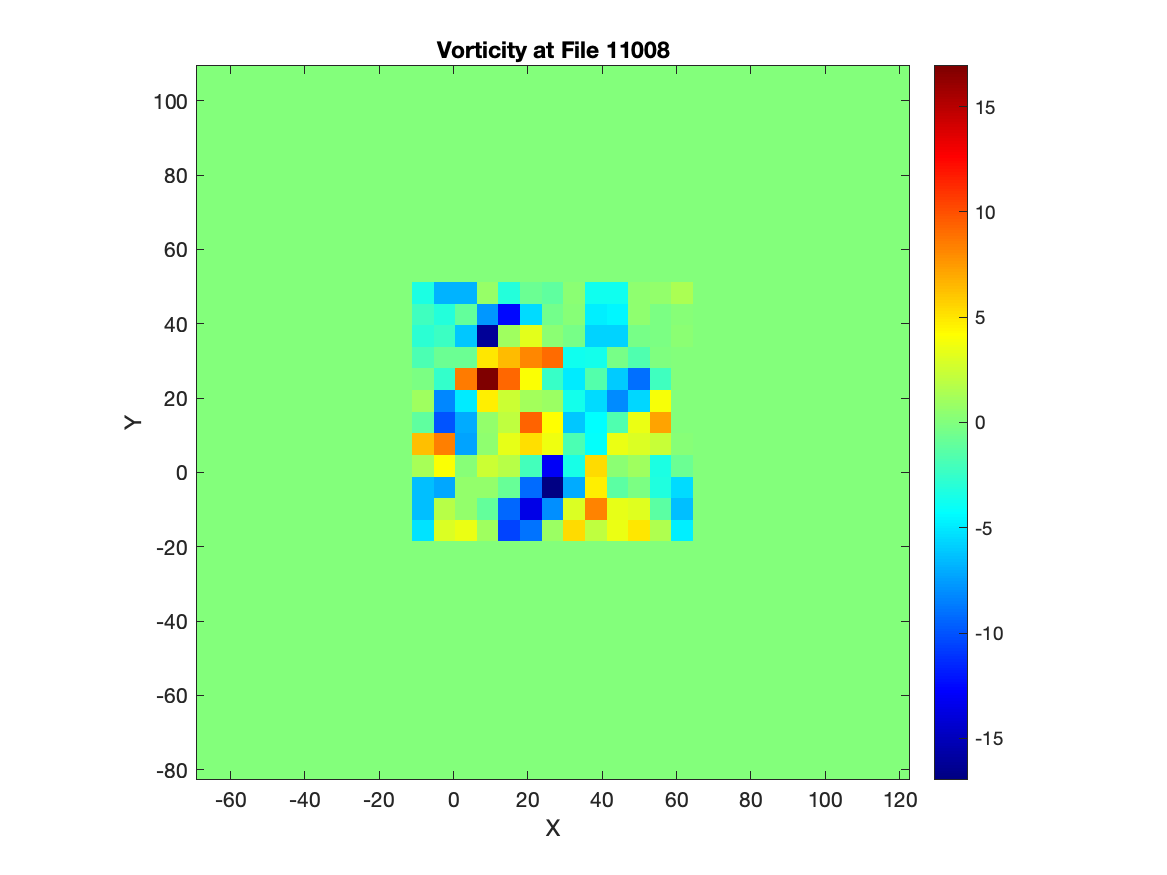}
    \caption{\label{fig:curl} Example of the vorticity of the velocity field at the same time as Supplementary Figure~\ref{fig:vfield}. The colour bar represents the magnitude of the vorticity.}
\end{figure}

\begin{figure}
    \centering
    \includegraphics[width=15cm]{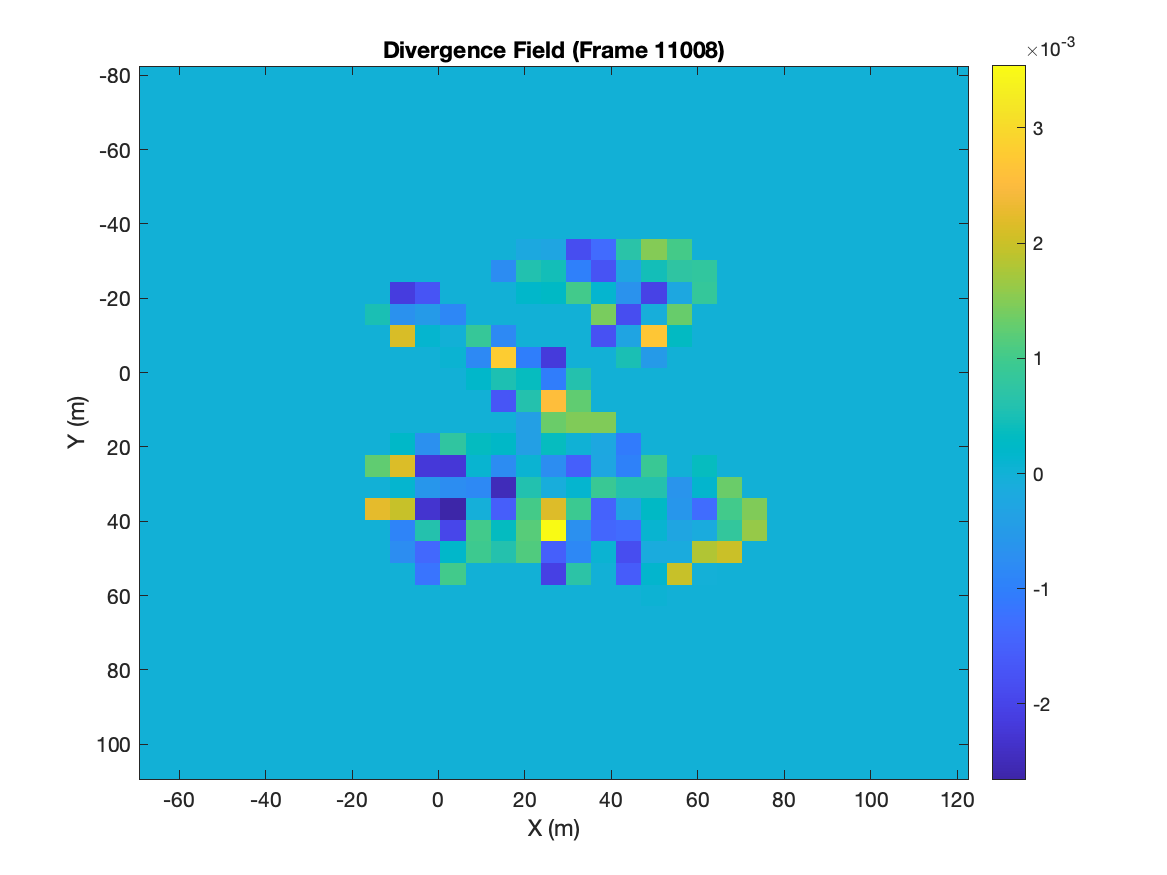}
    \caption{\label{fig:div} Example of the divergence of the velocity field at the same time as Supplementary Figures~\ref{fig:vfield} and~\ref{fig:curl}. The colour bar represents the magnitude of the divergence.}
\end{figure}

\begin{figure}
    \centering
    \includegraphics[width=15cm]{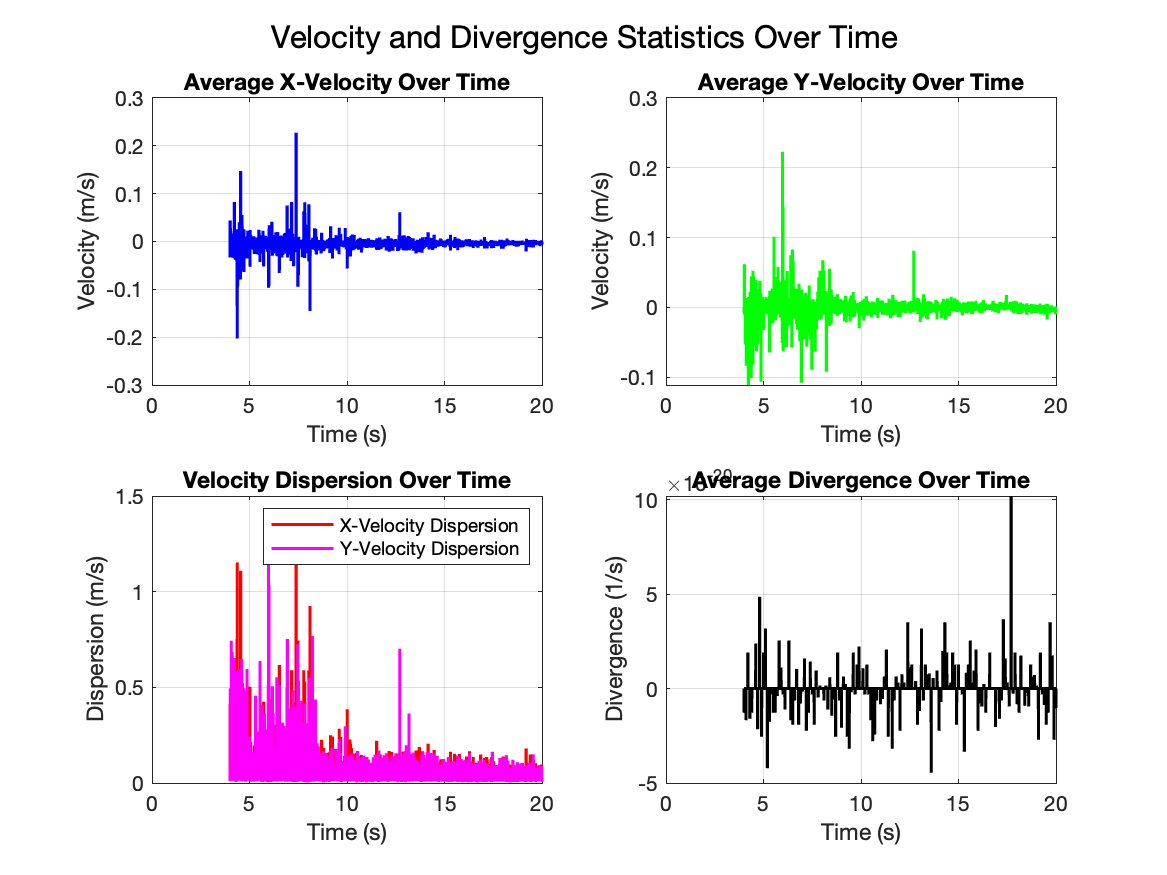}
    \caption{\label{fig:stats} Statistics of the velocity field over the duration of the measurement. Values cannot be calculated before second 4 because the particle seeding density is too low to derive a vector field, indicating an initially low dust-to-gas mass density ratio.}
\end{figure}

\begin{figure}
    \centering
    \includegraphics[width=15cm]{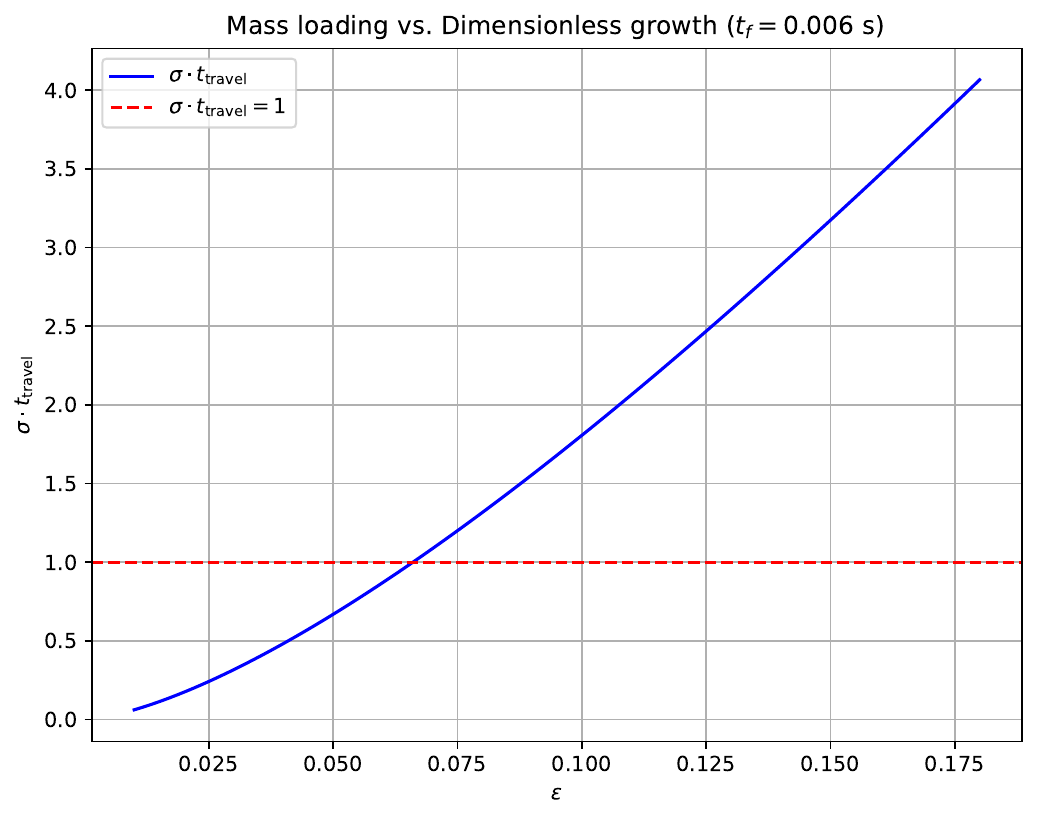}
    \caption{\label{fig:times} Comparison of the particle transport time to the expected instability growth rate for different values of the initial dust-to-gas mass density ratio, $\epsilon$. The horizontal dashed line shows when the growth rate and travel time to the measurement volume are equal.}
\end{figure}
\pagebreak
{\bf Supplementary References}\\
Larson Capelo, H.: Two-Phase Shear Flow Instability in Rarefied Gas.\\ https: //doi.org/10.5281/zenodo.10853364.

\end{document}